\documentclass[twocolumn,pre,aps,showpacs]{revtex4-1}
\usepackage{graphics,graphicx,epsfig}
\usepackage{amssymb}
\usepackage{textcomp}
\usepackage{color}
\begin{document}
\title{Substrate effects and diffusion dominated roughening in Cu$_2$O electrodeposition}
\author{I. S. Brandt${}^{1}$, V. C. Zoldan${}^{1}$, V. Stenger${}^{1}$, C. C. Pl\'a Cid${}^{1}$,
A. A. Pasa${}^{1,(a)}$, T. J. Oliveira${}^{2,(b)}$ and F. D. A. Aar\~ao Reis${}^{3,(c)}$
\footnote{a) andre.pasa@ufsc.br \\ b) tiago@ufv.br\\
c) reis@if.uff.br}}
\affiliation{${}^{1}$ Departamento de F\'isica, Universidade Federal de Santa Catarina,
88040-900, 476, Florian\'opolis, SC, Brazil\\
${}^{2}$ Departamento de F\'isica, Universidade Federal de Vi\c cosa, 36570-000, Vi\c cosa,
MG, Brazil\\
${}^{3}$ Instituto de F\'\i sica, Universidade Federal Fluminense, Avenida Litor\^anea s/n,
24210-340 Niter\'oi RJ, Brazil\\
}

\date{\today}

\begin{abstract}

Cuprous oxide (Cu$_2$O) films from 25 nm to 1500 nm were electrodeposited on n-Si(100)
and Ni/n-Si(100) substrates from aqueous solution at room temperature.
X-ray diffraction and transmission electron microscopy imaging show that the Cu$_2$O structure
and morphology is strongly affected by the substrate choice, with V shape and U shape
columnar growth on n-Si(100) and Ni/n-Si(100), respectively.
Atomic force microscopy reveals the presence of rounded grains at the surface in both cases.
Anomalous and normal roughening are observed in films grown on n-Si and Ni,
respectively, but estimates of scaling exponents are not conclusive.
On the other hand, the distributions of local heights, roughness, and extremal heights show
good agreement with those of the fourth order linear stochastic equation of Mullins and
Herring (MH). Thus, surface dynamics in both systems is dominated by
diffusion of adsorbed molecules, with no large scale effect of possible inhomogeneities
in mass flux from the solution or in reaction and adsorption rates.
In growth on n-Si substrates, the noise amplitude of the MH equation increases in time
as $t^{0.8}$, while the coefficient of the curvature-related term is time-independent.
Step edge energy barriers restrict the mass flux across grain boundaries,
thus a broad size distribution of initial grains leads to coarsening of the larger ones.
This explains their V shape in the thickest films and establishes a connection with
the anomalous roughening.
These effects are reduced in films grown on Ni/n-Si, which
initially have much larger grains with narrower size distributions and, consequently,
smaller fluctuations in coarse grained growth rates.
Thus, despite the relevance of electrochemical conditions for Cu$_2$O films to grow and
their influence on crystallographic orientation, large scale surface features are
determined by physical properties of the material and its interactions with the substrate,
with a universal microscopic dynamics similar to vapor deposition.

\end{abstract}

\maketitle

\section{Introduction}
\label{intro}

Cu$_2$O (cuprous oxide) is a p-type semiconductor that has recently attracted the attention
of experimental and theoretical groups due to its potential for application in metal base
transistors~\cite{Delatorre2006}, spintronic~\cite{Joseph2008}, photocathode for water
splitting~\cite{Morales-Guio2014}, electrochemical supercapacitors~\cite{Deng2013}
and catalysts~\cite{Liu2011}, and for light harvesting~\cite{Oba2005}.
The deposition of Cu$_2$O layers has been achieved by different techniques, such as pulsed
laser deposition~\cite{Chen2009}, magnetron sputtering~\cite{Deuermeier2011},
copper oxidation~\cite{Mittiga2006}, radical oxidation~\cite{Zang2013},
and electrodeposition~\cite{Golden1996}.
Electrodeposition is a very versatile technique, allowing to modify many structural, optical,
and electrical properties by the control of the deposition
parameters~\cite{Switzer2002,Brandt2014,Bijani2009,Han2009}.
Moreover, electrodeposition appears as a cost effective method to the preparation of metal
and semiconductor thin films for device applications.
When compared with physical deposition processes, such as sputtering, electrodeposition
has the particular feature of diffusive mass transport of species from the electrolyte to
the growing surface~\cite{Gamburg2011}.

A comprehensive understanding of the electrochemical growth mechanisms, taking into account
the mass transport at the electrode surface, plays a vital role to the production of
films with the desired properties.
One example is the technological requirement for mesoscopic layers to be used in photocathode
applications, in which the grain shape and the exposed crystalline faces need to be controlled
to improve stability and efficiency~\cite{Paracchino2012}.
On the other hand, the study of surface topography of deposited layers helps to discriminate
the basic mechanisms of growth dynamics from the particular features of each set of
physico-chemical deposition conditions~\cite{silvaSS2005,barabasi,krug,evansreview,Hua2011}.
A crucial condition may be the substrate, which affects the early stages of island formation
and growth and, possibly, the subsequent film dynamics.

Bearing in mind the above mentioned arguments, the present work aims to investigate the
structure of electrochemically grown Cu$_2$O films on top of a semiconducting and
monocrystalline substrate and on top of a metallic and polycrystalline substrate,
viz. n-Si(100) and Ni evaporated on n-Si(100).
Recently, our group showed that these two substrates are suitable for morphological,
structural and optical studies on Cu$_2$O~\cite{Brandt2014}.
Moreover, the Cu$_2$O/Ni structure may be employed on photochemical~\cite{Somasundaram2007}
and spin transport~\cite{Pallecchi2010} investigations.

Raman spectroscopy, X-ray diffraction (XRD), and transmission electron microscopy (TEM)
measurements are carried out to characterize film composition and structure.
Atomic force microscopy (AFM) images are used to study kinetic roughening of the film
surface, thus helping to disclose the main mechanisms of the growth dynamics.
Films directly grown on the semiconductor substrate have a V shape columnar structure,
preferential growth in $[100]$ direction, and show anomalous roughening, while those
grown on the Ni layer present an U shape columnar structure, change the dominant
crystallographic orientation during the growth, and have normal roughening.
A deeper analysis of surface morphology shows that the dominant mechanism in the 
Cu$_2$O growth is surface diffusion of adsorbed molecules, with uniform incoming flux.
Step edge energy barriers explain the anomaly of the former films, connecting this
feature to the grain shape.
The universal coarse-grained growth dynamics is similar to what is observed in many vapor
deposited films, despite the relevance of electrochemical conditions to local reaction and
adsorption processes.
On the other hand, the interaction between the film and the substrate,
which is frequently restricted to island nucleation and growth, here is shown to have
drastic consequences to the structure of the thickest Cu$_2$O films.

The rest of this work is organized as follows. In Sec. \ref{Secexper}, the experimental
procedure for growth and characterization of the Cu$_2$O films is described.
In Sec. \ref{Secresults}, the experimental results are presented and analyzed in the
light of kinetic roughening concepts. Section \ref{Secconclusion} presents a discussion
of the results and our conclusions.

\section{Experimental}
\label{Secexper}

Cu$_2$O films were deposited on (100) monocrystalline n-type silicon (resistivities in
the range of 6 - 9 \textohm$\cdot$cm) without and with a cap layer of 50 nm of evaporated Ni.
Before Cu$_2$O electrodeposition or Ni deposition, the silicon substrates were immersed
in HF 5\% for 20s to remove silicon native oxide from the surface.
The roughnesses of the n-Si substrate and of the Ni layer surface are respectively
$0.2$ nm and $1.3$ nm.
Ni thin films deposited on Si(100) had preferential growth in the $[111]$ direction,
which was checked by XRD measurements.

The electrochemical experiments were conducted in a conventional three electrode cell
connected to a potentiostat Autolab PGSTAT30. The counter electrode was a Pt foil and the
reference electrode was a Saturated Calomel Electrode (SCE). The electrolyte, prepared from
analytical grade reagents and deionized water (resistivity of 18 M \textohm$\cdot$cm),
contains $0.4$ M CuSO$_4$ and $3.0$ M lactic acid, with the pH adjusted to $10.00$ by
adding a solution of $5.0$ M NaOH~\cite{Golden1996}.

The deposits were obtained via potentiostatic experiments.
Samples were deposited at 25 \textcelsius~ for a deposition potential of $-0.5$ V
\textit{vs} SCE, where the only varied parameter was the thickness $H$ of the samples
for the 2 different substrates, n-Si(100) and Ni/n-Si(100).

The efficiency of the Cu$_2$O electrodeposition process was checked by Rutherford
backscattering in a previous work and values of 90\% were obtained~\cite{Delatorre2009}.
The samples were characterized by various techniques including Raman spectroscopy
(inVia, Renishaw), XRD (XPERT, Philips), TEM (JEM-2100, JEOL) and AFM
(Pico-SPM, Molecular Imaging Corporation).
Raman spectra were obtained from $514.5$ nm wavelength excitation (argon ion laser).
The AFM images were acquired in contact mode in air at room temperature and the values
of the root mean-square deviation of the local surface height, $W(l,H)$, are representative
of at least three images of different regions of the surface. Three different regions 
for sample were measured with scan sizes of $1 \times 1$, $2 \times 2$ and $5 \times 5\mu m$ 
for each region and number of pixels per line ranging from 256 to 1024, respectively. 
From these measurements we have concluded that for $2 \times 2 \mu m$ image size the $W(l,H)$ 
reaches the saturation value. In addition, before starting measurements on Cu$_2$O samples, 
the  performance of the AFM tip was checked by imaging Au thin films with known surface.

\section{Results}
\label{Secresults}

\subsection{Characterization of film composition and structure}
\label{ResCharact}

The copper oxide phase formed in the electrodeposited thin films was analyzed by Raman
spectroscopy and XRD.
The measurements were done on layers with thickness between 150 and 1500 nm deposited
on n-Si(100) and Ni/n-Si(100).

\begin{figure}[h]
\includegraphics[width=8.5cm]{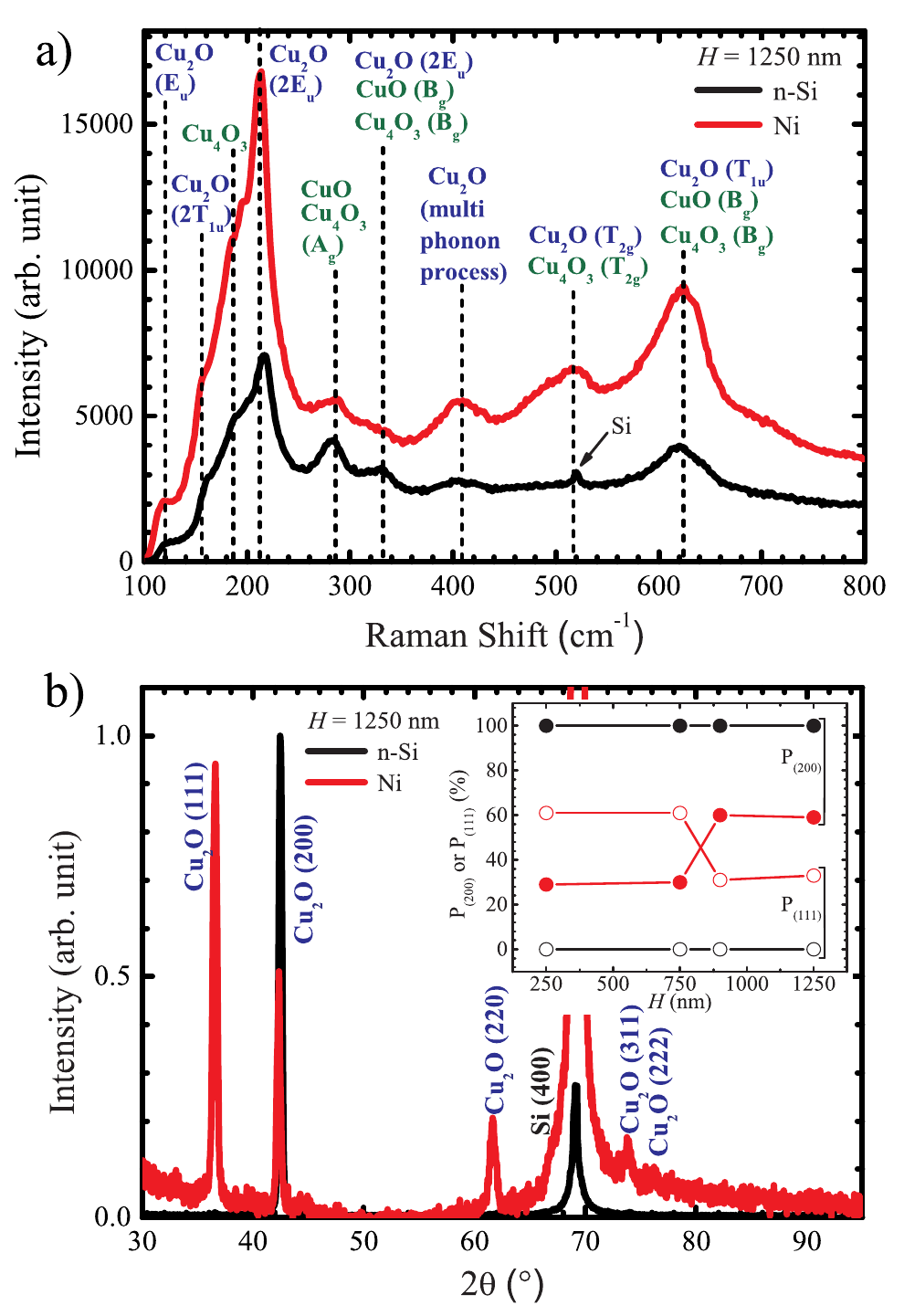}
\caption{(Color online) a) Raman spectroscopy and b) XRD results for Cu$_2$O films of 1250 nm electrodeposited on n-Si and Ni substrates. For each Raman peak are indicated the possible related crystalline phases and, between parentheses, the Raman vibrational mode. XRD characterization was carried out in Bragg-Brentano configuration. The inset in b) shows the Cu$_2$O portion growing in $[100]$ and $[111]$ directions as a function of the film thickness.}
\label{fig1}
\end{figure}

Raman spectroscopy results of films with thickness $1250$ nm are shown in Fig. \ref{fig1}a.
The predominance of Cu$_2$O phase is evident.
Contributions of Cu$_4$O$_3$ and possibly CuO phases are also observed.
The peaks in Fig. \ref{fig1}a were addressed to the respective phase and Raman vibration
mode based on Ref. \cite{Meyer2012}. Despite the observation of Cu$_4$O$_3$ and CuO Raman
peaks, these phases are not present in XRD Bragg-Brentano patterns of the same films,
shown in Fig. \ref{fig1}b, probably due to the higher surface sensitivity of Raman measurements. 
Cu$_4$O$_3$ and CuO phases are likely to be formed by oxidation of Cu$_2$O top layers.
Since the layers are mostly of Cu$_2$O, this work will be concerned to the description
of the electrochemical growth of this phase.

Fig. \ref{fig1}b also shows the XRD patterns of Cu$_2$O films with 1250 nm of thickness
deposited on Ni/n-Si(100). The peaks are also in the expected positions for the Cu$_2$O crystal.
Cu$_2$O/n-Si(100) samples display strong preferential growth in $[100]$ direction,
following the orientation of the substrate. On the other hand, Cu$_2$O film grown on top
of Ni/n-Si(100) substrate is composed by grains oriented in $[100]$, $[110]$, $[111]$,
and $[311]$ directions.

The evolution of Cu$_2$O growth on Ni/n-Si(100) is explained by Ref. \cite{Brandt2014}
as follows: (\textit{i}) the Cu$_2$O films show initial $[111]$ growth due to a better
coupling of the Cu$_2$O(111)/Ni(111) interface compared to the Cu$_2$O(100)/Ni(111) one;
(\textit{ii}) at electrolyte pH of $10.00$ the $[100]$ growth is favored, the initial
$[111]$ growth starts to be suppressed by the $[100]$ one and at a specific thickness
a major part of the Cu$_2$O deposit will be composed by $[100]$ grains.

These growth steps are confirmed by the inset of Fig. \ref{fig1}b, which shows the
fraction of the material growing in $[100]$ and $[111]$ directions as a function of
film thickness for both substrates.
While Cu$_2$O on n-Si(100) is, independently of film thickness, oriented in $[100]$ direction,
Cu$_2$O on Ni/n-Si(100) grows up to 750 nm with $\sim 60$\% of its grains oriented in $[111]$
direction, but only $\sim 30$\% remain with this orientation in larger thicknesses.
The initial growth of Cu$_2$O on Ni(111) is thermodynamically controlled by the
Cu$_2$O(111)/Ni(111) coupling, likewise as previously reported for Cu$_2$O(111)/Au(111)
and Cu$_2$O(100)/Au(100) interfaces~\cite{Switzer2002}.
However, the crystallographic orientation transition at 750 nm is related to the oxygen
concentration in electrolyte at pH $10.00$, which favors the $[100]$ growth~\cite{Brandt2014}.

\begin{figure}[t]
\includegraphics[width=8.5cm]{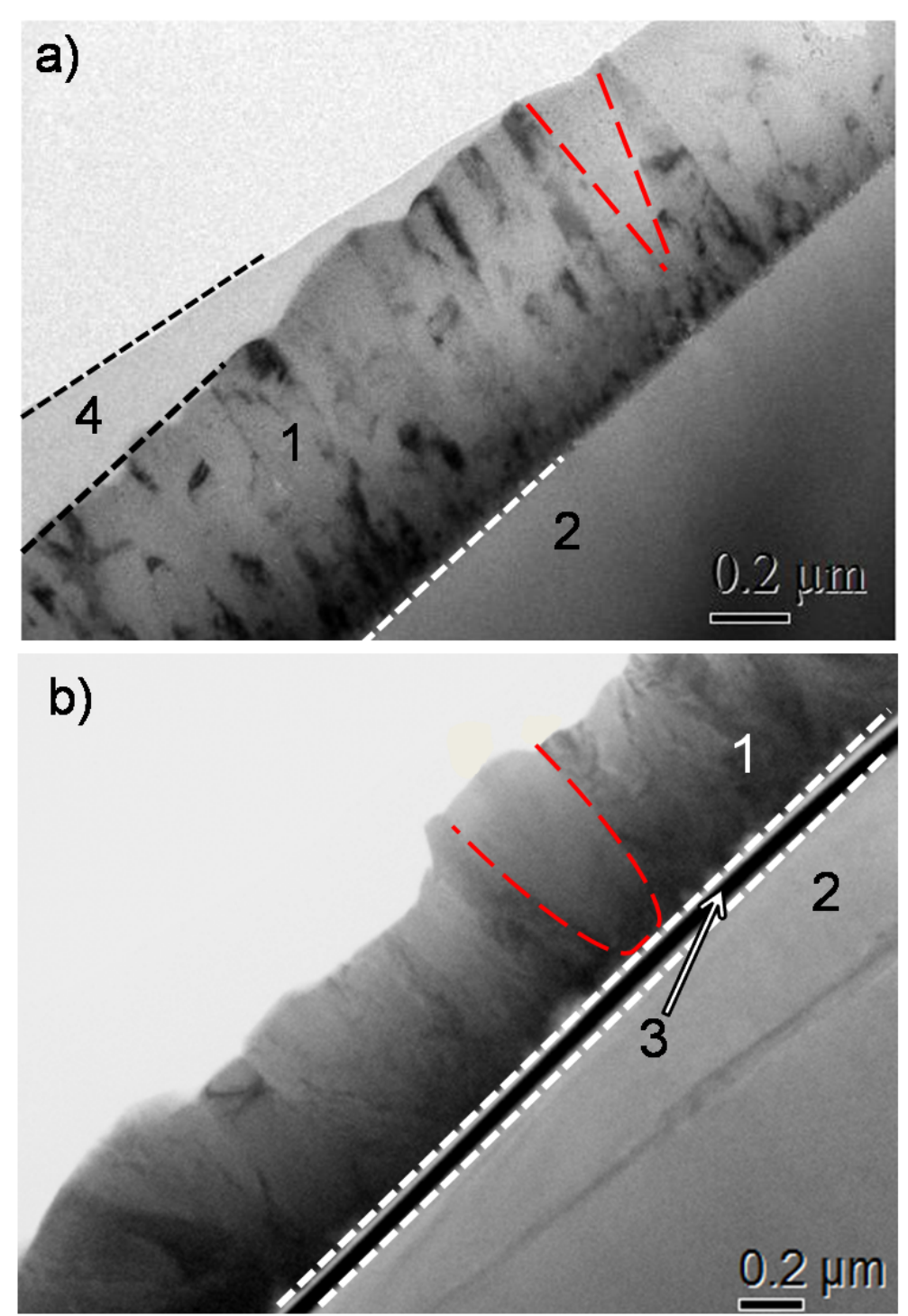}
\caption{(Color online) Cross section TEM images of Cu$_2$O thin films electrodeposited
respectively on a) n-Si and b) Ni/n-Si. Different layers in each image are labeled as follows:
1 Cu$_2$O film, 2 n-Si substrate, 3 Ni film (substrate) and 4 the glue used on sample
preparation. In a) and b) discontinuous red lines highlight the V shape and U shape
of the Cu$_2$O columns, respectively.}
\label{fig2}
\end{figure}

The structure of Cu$_2$O films was checked by cross section TEM images.
Figure \ref{fig2}a shows an image obtained from electrodeposited Cu$_2$O film on n-Si(100).
A high density columnar microstructure without voids is observed.
These columns have a V shape as stressed by discontinuous red lines, exhibiting a grain
width increase as a function of film thickness.
On the other hand, as seen in Fig. \ref{fig2}b, Cu$_2$O films deposited on the Ni layer
under same conditions (solution, temperature and deposition potential) displayed a columnar
microstructure with U shape, as indicated in Fig. \ref{fig2}b, and grain width roughly
constant as a function of growth time.

\begin{figure}[t]
\includegraphics[width=8.5cm]{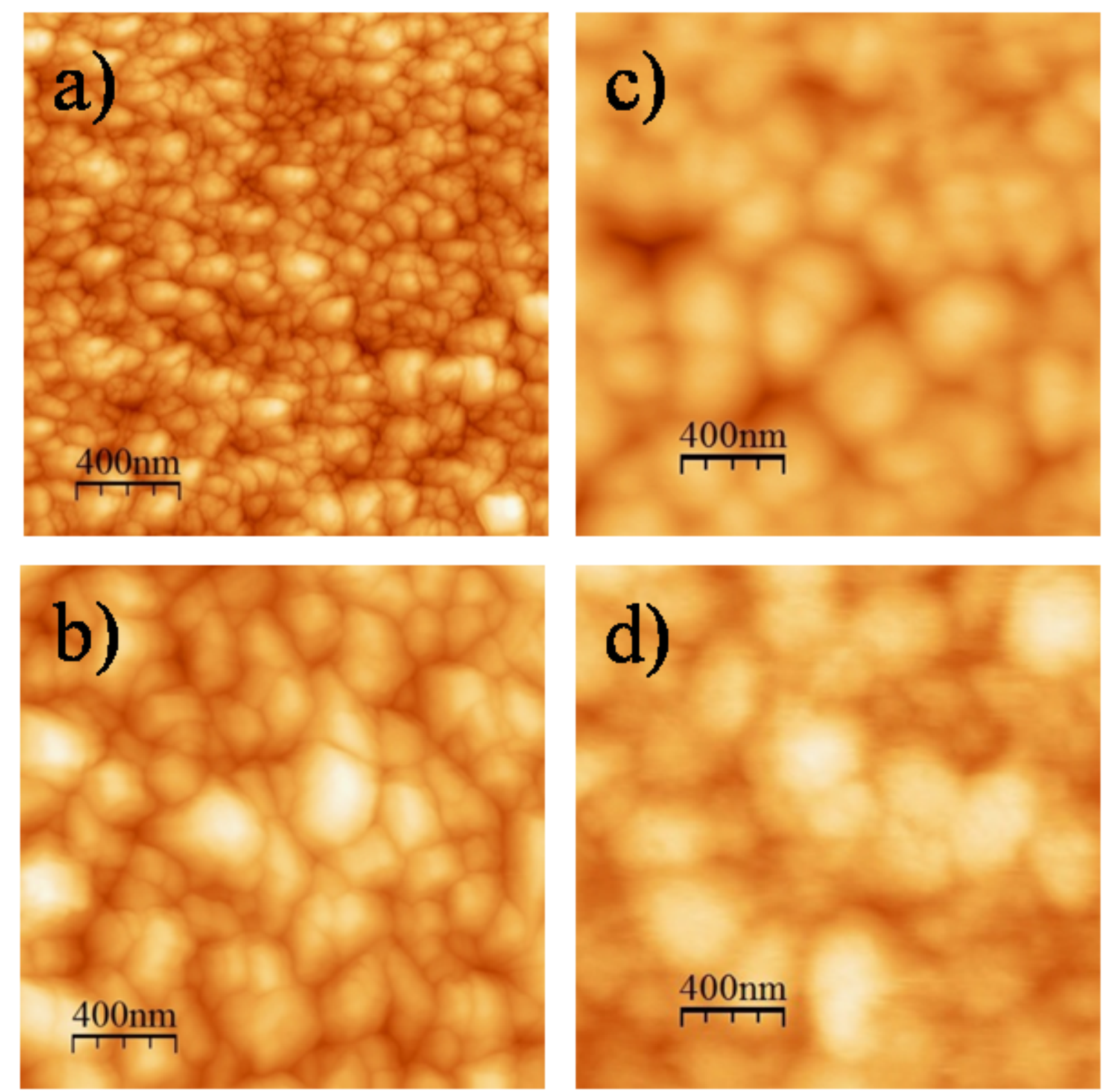}
\caption{(Color online) $2 \mu m \times 2 \mu m$ AFM images of the surface morphology
of Cu$_2$O layers electrodeposited on: a), b) n-Si; and c), d) Ni/n-Si.
Upper row is for thickness of 250 nm, while lower row is for 1500 nm.
Bright areas are elevations and dark ones are deep regions.}
\label{fig3}
\end{figure}

Figure \ref{fig3} shows $2 \times 2 \mu m^2$ AFM images of the films with thicknesses
of 250 and 1500 nm.
A common feature in both substrates is that the film surface presents a granular aspect.
In films deposited on n-Si, there is a remarkable increase of the grain size from the
thinnest to the thickest film. The columnar structure with V shape shown in TEM images
and this increase in grain size suggest a coarsening process, with part of the initial
columns growing and enlarging up to 1500 nm at the expenses of neighboring ones.
On the other hand, a mild increase of grain size is observed in the films grown on Ni/n-Si,
which is in agreement with TEM images showing columns in U shape.

\subsection{Dynamic scaling of surface roughness}
\label{ResDynamic}

The surface roughness $W$ is the most used quantity to characterize height fluctuations.
For a given surface under a process of kinetic roughening, $W$ depends on the lengthscale
in which fluctuations are measured. Moreover, given a scanning box with lateral size $l$,
the fluctuation in film height inside the box also depends on the box position.

For these reasons, the box roughness is defined as the root-mean-square (rms) fluctuation
of the film height ($h$) inside a box at a given position, i.e,
$w = \sqrt{\overline{h^2} - {\overline{h}}^2}$.
The average local roughness $W(l,H) = \left\langle w\right\rangle$ of a film of thickness
$H$ in the lengthscale $l$ is the average of the box roughness $w$ among all possible box
positions. $W(l,H)$ is hereafter simply called roughness.

\begin{figure}[t]
\centering
\includegraphics[width=8.5cm]{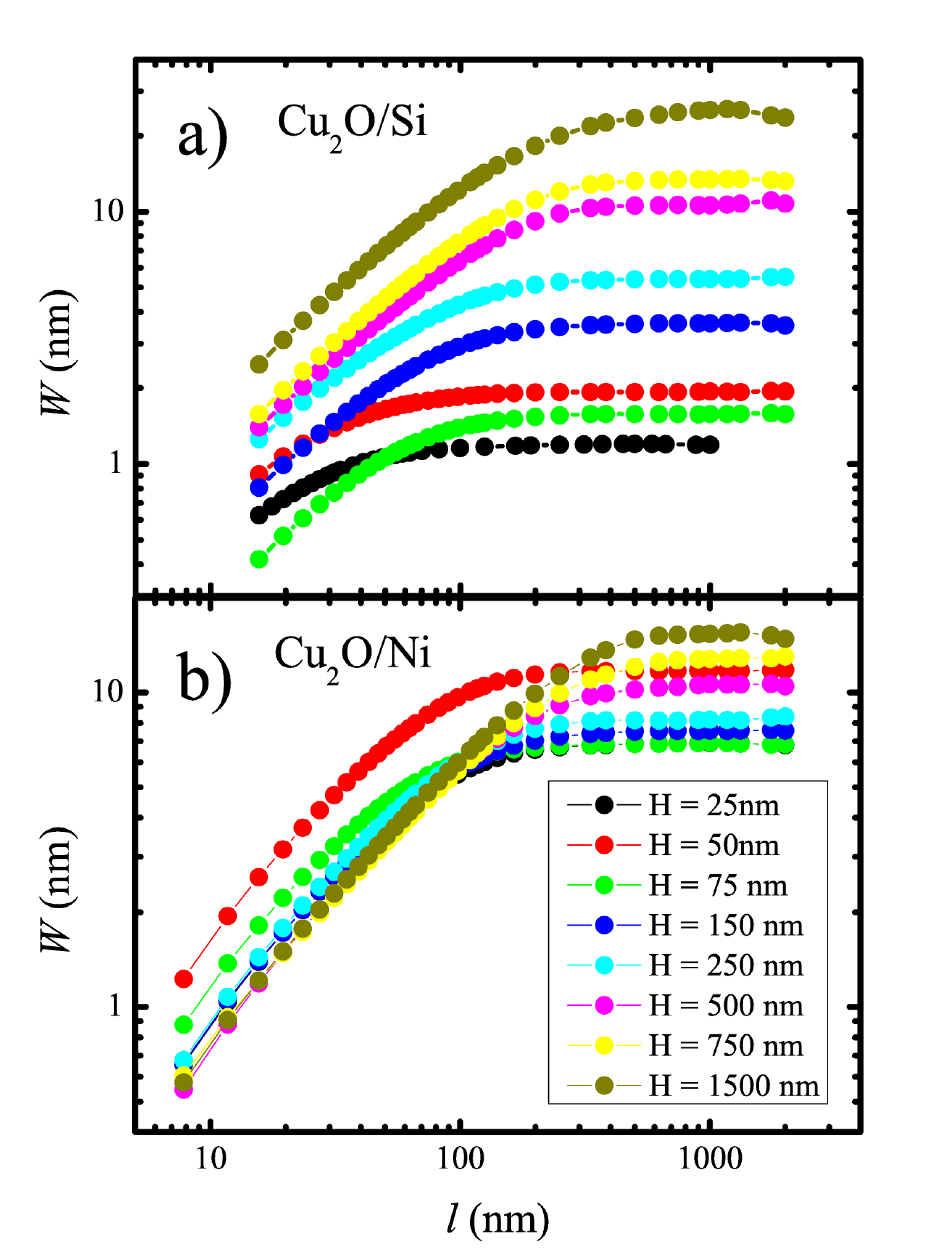}
\caption{(Color online) Roughness $W(l,H)$ as a function of the box size $l$ measured from AFM images of a series of electrodeposited Cu$_2$O films for thickness $H$ from 25 to 1500 nm. In a) and b) the substrate is n-Si and Ni/n-Si, respectively.}
\label{fig4}
\end{figure}

Using the heights from AFM images, $W(l,H)$ is calculated for Cu$_2$O layers grown on
both substrates. The results are shown in Figs. \ref{fig4}a and \ref{fig4}b for a range
of thicknesses $H$ varying from 25 to 1500 nm. The structural changes induced by substrates
n-Si and Ni/n-Si are reflected in topographic differences in the growing surface of the
films, confirming the visual inspection of AFM images.

The roughness $W(l,H)$ of films grown on Ni/n-Si show normal dynamic scaling because
it is approximately time-independent for small box size $l$ and saturates at values
increasing with the thickness $H$. Deviations for the thinner samples (25 to 75 nm) are
characteristic of the initial island growth and coalescence, indicating that an extended
film is formed between 75 nm and 150 nm of thickness. For this reason, only results for
$H \geqslant 150$ nm will be analyzed below.

\begin{figure*}[t]
\centering
\includegraphics*[width=15.cm]{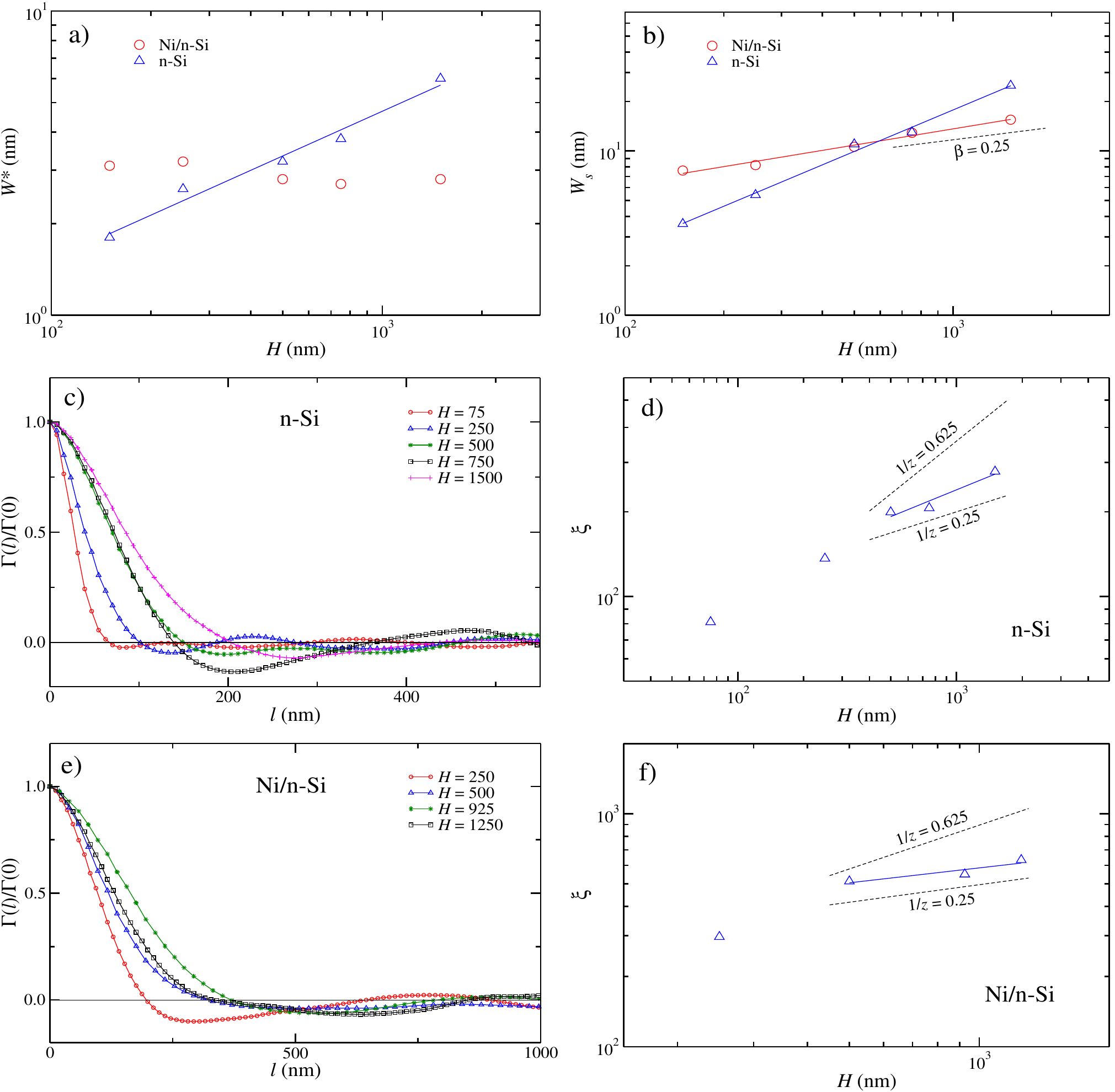}
\caption{(Color online) a) Local roughness for box size $l = 40$ nm ($W^{*}$) and
b) global roughness ($W_s$) as functions of the film thickness $H$.
Rescaled correlation functions $\Gamma(l)/\Gamma(0)$ against the length $l$ for
c) n-Si and e) Ni/n-Si substrates. Their first minima (the correlation length $\xi$) are
shown in d) and f) as functions of $H$. Continuous lines are power law fits of the
data and dashed lines have the slopes indicated.}
\label{fig5}
\end{figure*}

On the other hand, the oxide layers grown on n-Si have anomalous scaling, since the
roughness in small boxes shift to larger values with increasing
thicknesses~\cite{lopez,ramasco,huo,lafouresse}. The presence of a single dominant
orientation during the growth (shown by XRD) and the coarsening of growing columns from
the smallest to the largest thicknesses (suggested by TEM and AFM images) indicate that
the scaling anomaly is an intrinsic feature of the surface dynamics in this system.

The data presented in Fig. \ref{fig4} for the thickest films with $l \lesssim 100$ nm 
were fit to the relation $W \sim l^{\alpha_l}$.
It gives estimates of the local roughness exponents $\alpha_l = 0.90$ and $0.91$ in n-Si
and Ni/n-Si substrates, respectively.
These values are consistent with the interpretation of granular effects in Refs.
\cite{tiago2007} and \cite{tiago2011}, which predict an exponent near 1 if the maximal
box size is smaller than the grain size.
Inspection of the AFM images (Fig. \ref{fig3}) supports this hypothesis.
These estimates of $\alpha_l$ must not be interpreted as true exponents of dynamic scaling,
but as exponents representing the grain shapes~\cite{tiago2011}.

The local roughness $W^{*} \equiv W(l^{*},H)$ is measured for a fixed small box size
$l^{*} = 40$ nm. It is shown in Fig. \ref{fig5}a as a function of the thickness $H$.
For films deposited on Ni, $W^{*}$ fluctuates around a constant value as the thickness
increases, from small values to the largest ones.
This is consistent with normal scaling. On the other hand, for films deposited
on n-Si, $W^{*}$ shows a power law increase from small thicknesses ($H=150$ nm) to the
largest ones ($H=1500$ nm). This is the main evidence of anomalous scaling in these films.
The fit of the data in Fig. \ref{fig5}a by the scaling relation
$W^{*}\sim H^\kappa$ \cite{lopez}
gives an anomaly exponent $\kappa = 0.49$ ($\kappa$ is also called local
slope exponent in theoretical works and local growth exponent $\beta_{loc}$ in experimental
works~\cite{huo,lafouresse}).

The saturation values of the roughness in Fig. \ref{fig4} give the global roughness $W_s$,
which characterize fluctuations of the whole surface at a given thickness.
In Fig. \ref{fig5}b, $W_s$ is shown as a function of $H$ for both films.
The fits to the scaling relation $W_s\sim H^\beta$
for $H \geqslant 150$ give the growth exponents $\beta = 0.84$ and
$\beta = 0.33$ for films deposited on n-Si and Ni/n-Si substrates, respectively.

Since the growth on Ni/n-Si substrates has normal scaling, we search for a model of
local stochastic growth equation, which may disclose the main mechanisms of the surface
dynamics~\cite{barabasi,krug}.
The above estimate of $\beta$ is larger than the values provided by the equations
of second and fourth orders, which are in the range $[0, 0.25]$~\cite{barabasi}.
However, Fig. \ref{fig5}b shows that a slope close to $\beta = 0.25$ cannot be discarded
for large thicknesses. This is close to the exponents of the Kardar-Parisi-Zhang (KPZ)
class~\cite{kpz}, $\beta \approx 0.24$~\cite{fabio2004,kelling}, and of the
Mullins-Herring (MH) class~\cite{mh}, $\beta = 0.25$.

Information on surface correlations can be obtained from the spatial correlation function
$\Gamma\left( l,t\right) \equiv \left\langle \tilde{h}(x+l,t) \tilde{h}(x,t) \right\rangle$
at time $t$, with $\tilde{h} \equiv h - \left\langle h \right\rangle$.
The rescaled functions $\Gamma\left( l,t\right)$ of the oxide films respectively grown
on n-Si and Ni/n-Si substrates are shown in Figs. \ref{fig5}c and \ref{fig5}e as a function
of the distance $l$, for several thicknesses.
A reliable estimate of the correlation length $\xi$ is the minima of
$\Gamma\left( l,t\right)$~\cite{Almeida2014,siniscalco}, which is shown in Figs. \ref{fig5}d and \ref{fig5}f as a function of the thickness $H$.

The estimates of $\xi$ are close to those obtained by intersection of straight line fits
of the roughness in the growth and saturation regimes, which is a usual method in
the analysis of experimental data.
Ref. \protect\cite{zoldan2005} shows that this method leads to estimates of $\xi$
of the same order of the average grain size estimated by inspection of
microscopy images~\cite{zoldan2005}.
However, an accurate calculation of the average grain size is more complicated
\cite{siniscalco} and not performed here.

The scaling $\xi \sim t^{1/z} \sim H^{1/z}$ is expected for large thicknesses~\cite{barabasi,krug}.
Linear fits of the data for $H \geqslant 500$ nm give estimates $1/z = 0.32$ and $0.21$
(dynamic exponents $z = 3.1$ and $4.7$) for growth on n-Si and Ni/n-Si substrates,
respectively.
These exponents are very different from the KPZ value $z \approx 1.6$~\cite{fabio2004,colaiori}.
On the other hand, the Figs. \ref{fig5}d and \ref{fig5}f show that the slopes
of the $\log{\xi} \times \log{H}$ plots for the largest thicknesses are close to the MH
value $1/z =0.25$ (the corresponding slope of KPZ scaling $1/z = 0.625$ is also shown
for comparison).

The MH equation is a model for growth dominated by surface diffusion of adsorbed
species~\cite{mh}:
\begin{equation}
 \frac{\partial h}{\partial t} = - K \nabla^{4} h + \eta(\vec{x},t) ,
\label{eqMH}
\end{equation}
where $h\left( \vec{x},t\right)$ is the interface height at substrate position $\vec{x}$
and time $t$, $K$ is a constant, and $\eta$ is a Gaussian (nonconservative) noise
with covariance amplitude $D$~\cite{barabasi,mh}.
A constant external flux was omitted from Eq. (\ref{eqMH}), which corresponds to a spatially
uniform and time-independent adsorption rate. Thus, the local growth rate of the MH model
is not affected by diffusion in solution, for instance due to the presence of diffusive
layers, shadowing effects etc.

On the other hand, the KPZ equation is a model for interface growth dominated
by surface tension and by a nonlinear effect of the local slope~\cite{kpz}:
\begin{equation}
\frac{\partial h}{\partial t} = \nu{\nabla}^2 h + \frac{\lambda}{2}
{\left( \nabla h\right) }^2 + \eta (\vec{x},t) ,
\label{kpz}
\end{equation}
where $\nu$ and $\lambda$ are constants. 
Distinguishing these possibilities is an important step to understand the microscopic
growth dynamics of the oxide films. The KPZ model also assumes spatially uniform and
time-independent adsorption rate, with no effect of diffusion in solution to the local
growth rate.

\subsection{Scaling of distributions}
\label{distributions}

We now turn to the analysis of distributions of heights and of local quantities,
which was recently shown to be a more powerful tool to the study of kinetic roughening
of thin films~\cite{Almeida2014,healy2014,Almeida2015}.

\begin{figure*}[t]
\centering
\includegraphics*[width=15.cm]{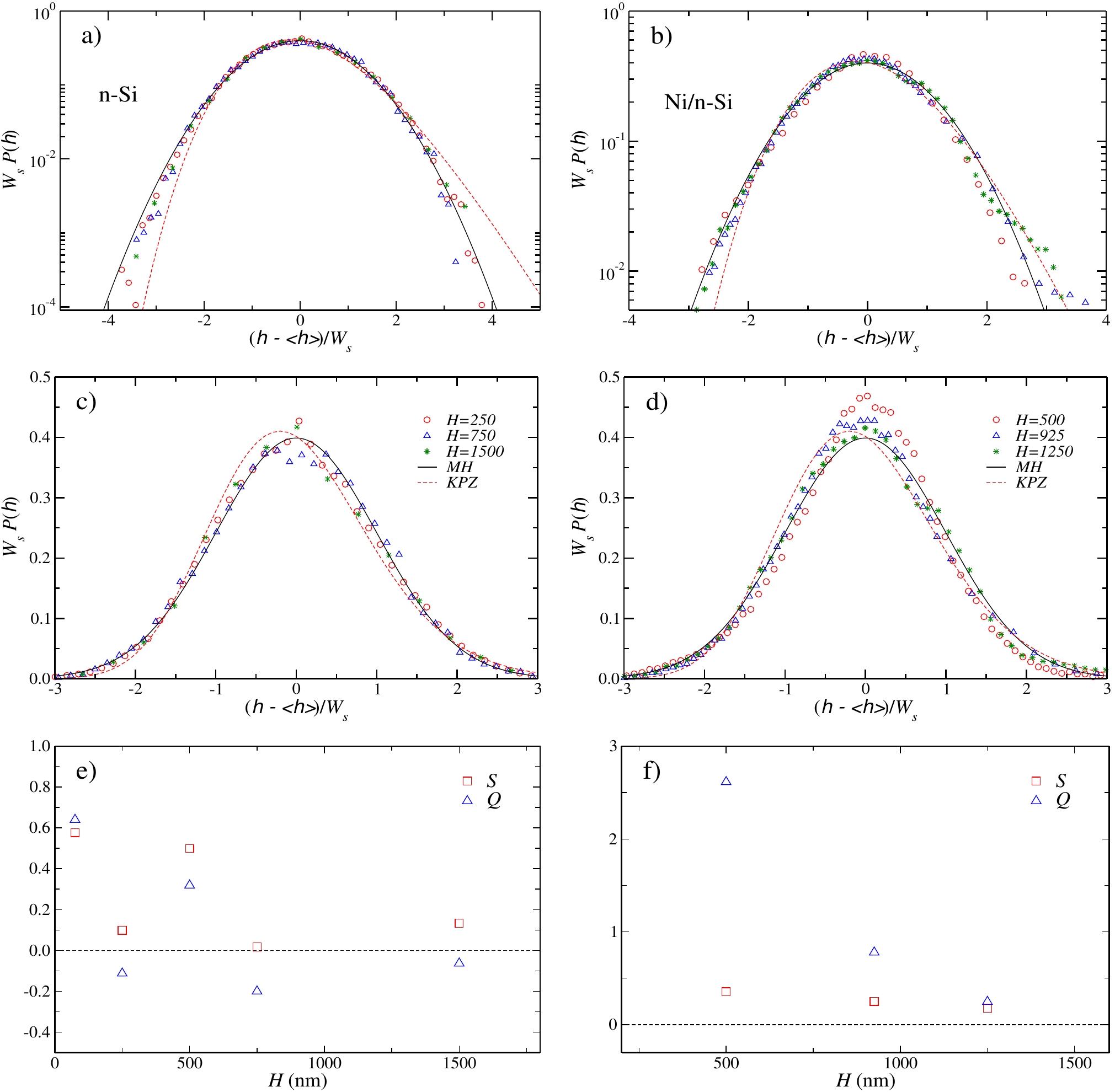}
\caption{(Color online) Scaled height distributions for a) and c) n-Si, and b) and d) Ni/n-Si substrates.
In a) and b) the data are presented in semi-log scale and in c) and d) in linear scale.
The skewness $S$ and kurtosis $Q$ of the height distributions as functions of the
thickness $H$ are shown in e) and f) for n-Si and Ni/n-Si, respectively.}
\label{fig6}
\end{figure*}

The height distribution $P(h)$ is defined as the probability density of the height $h$,
so that $P(h)dh$ gives the probability of finding a height in the range $[h, h + dh]$.
Figures \ref{fig6}a and \ref{fig6}c, and \ref{fig6}b and \ref{fig6}d show the scaled height distributions of the oxide
films with large thicknesses, respectively grown on n-Si and Ni/n-Si substrates.
They are compared with distributions of the KPZ and MH classes in the growth regimes.
The KPZ distribution was numerically calculated in Refs. \cite{halpin2012} and
\cite{tiagorapid2013}. For the MH class, the distribution is Gaussian, which was
confirmed by direct integration of the MH equation [Eq. \ref{eqMH}] and by simulation
of the large curvature model (LCM)~\cite{lcm}.

The experimental data in Figs. \ref{fig6}a, \ref{fig6}b, \ref{fig6}c and \ref{fig6}d are closer to the
Gaussian/MH curve. The symmetric shape clearly contrasts with the asymmetry of the KPZ
curve. Indeed, the skewness $S$ and kurtosis $Q$ of the $P(h)$ distributions, shown in
Figs. \ref{fig6}e and \ref{fig6}f, drops to zero as the thickness
increases (consistently with Gaussian/MH), while the asymptotic KPZ values are
$S \approx 0.43$ and $Q \approx 0.34$~\cite{halpin2012,tiagorapid2013}.

Another important conclusion emerges from Figs. \ref{fig6}a and \ref{fig6}b:
since the scaled height distributions are the same for films deposited on n-Si and
on Ni/n-Si, the main physico-chemical processes responsible for their roughening
are the same.

\begin{figure*}[t]
\includegraphics*[width=15.cm]{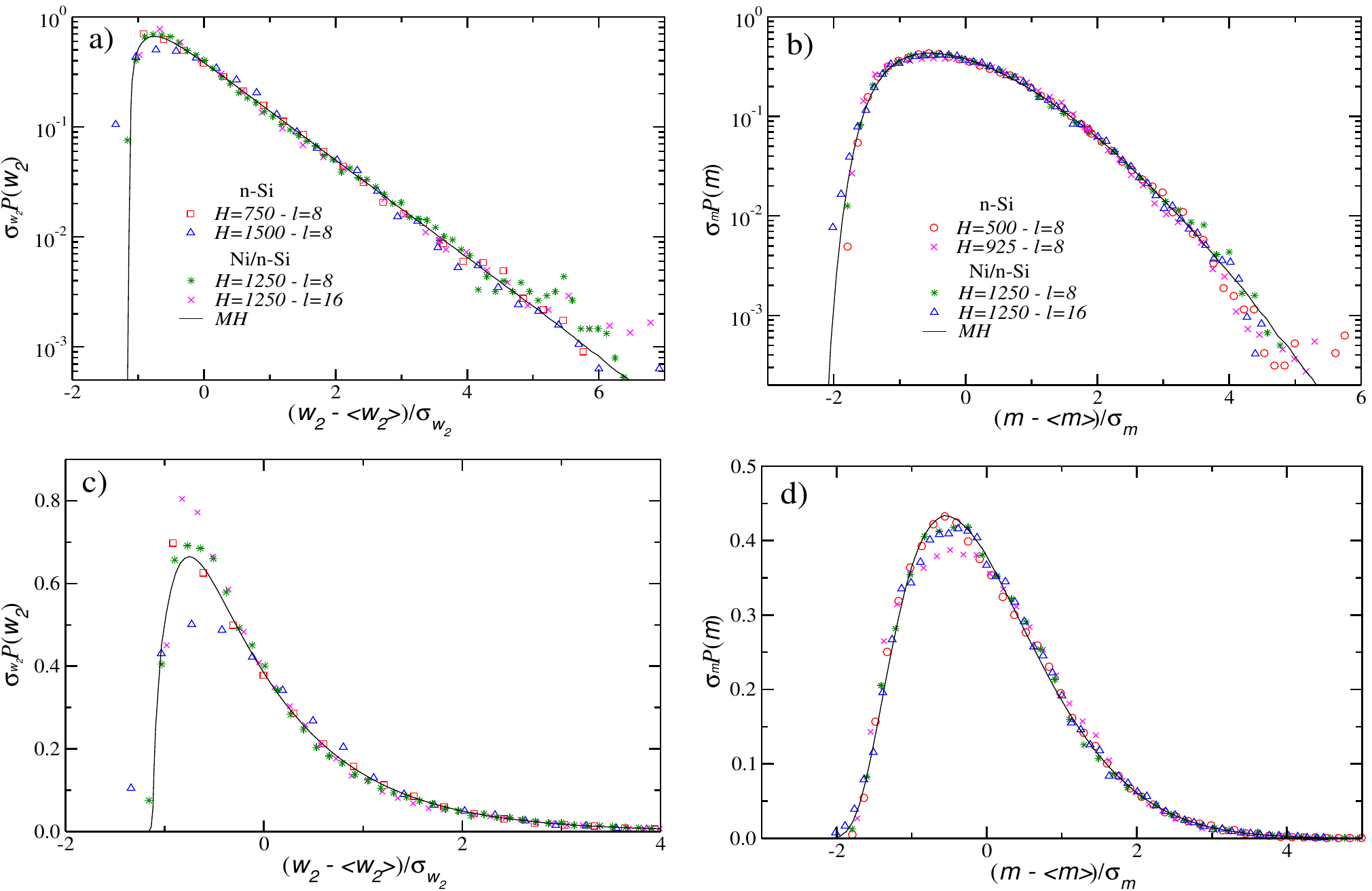}
\caption{(Color online) Scaled a) squared roughness and b) maximal relative height
distributions for both substrates and several thickness $H$ and box sizes $l$
(measured in pixels of $256 \times 256$ images). The same data are shown in linear scale in c) and d).}
\label{fig7}
\end{figure*}

The distribution $P(w_2)$ of the squared box roughness $w_2 = w^2$ is defined so that
$P(w_2)dw_2$ is the probability that the squared roughness in a given box is in the
range $[w_2, w_2+dw_2]$~\cite{racz1994,antal2002}.
Figures \ref{fig7}a and \ref{fig7}c compares the scaled roughness distributions of the oxide films
grown on both substrates and that of the MH class in the growth regime ($\sigma_{w_2}$
is the rms fluctuation of $w_2$).
The size of the box in which $w_2$ is measured ranges from 100 nm to 300 nm.
The MH curve is calculated by integration of the MH equation in the growth regime.

In Figs. \ref{fig7}a and \ref{fig7}c, the collapse of the scaled distributions for films grown in
different substrates is clear, which is additional support to the proposal that their
roughening is governed by the same physico-chemical processes (despite the very different
structures of those films). Good agreement with the MH curve reinforces the proposal that
diffusion of adsorbed species is the main mechanism of the surface dynamics.

We also analyzed the distribution $P(m)$ of the maximal relative height
$m \equiv h_m- \left\langle h\right\rangle $ in the growth regime.
Here, $h_m$ and $\left\langle h\right\rangle$ are, respectively, the maximal and the
average heights measured inside each box position.
The distributions for both films are shown in Figs. \ref{fig7}b and \ref{fig7}d, showing good data collapse
among them and with the MH curve.

\begin{figure*}[t]
\includegraphics[width=16.0cm]{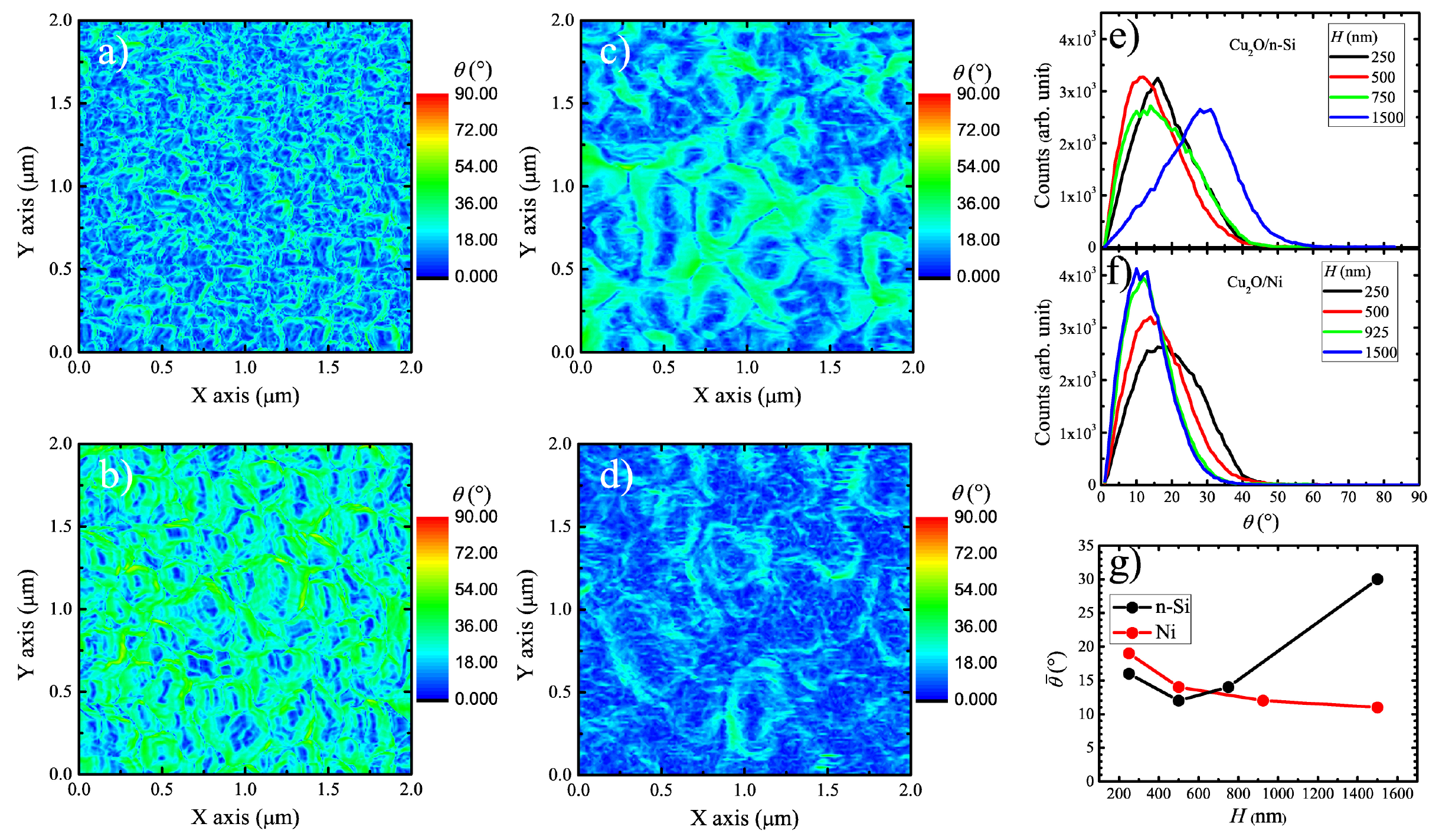}
\caption{(Color online) Slope images of Cu$_2$O films on n-Si, a) and b), and Ni/n-Si, c) and d), evaluated from height images shown in Fig. \ref{fig3}. In the first and second row Cu$_2$O films have 250 and 1500 nm of thickness, respectively. The color bar indicates the surface slope $\theta$, which can vary from 0 to 90\textdegree. Slope histograms for e) Cu$_2$O/n-Si and f) Cu$_2$O/Ni structures, and g) mean surface slope as a function of the film thickness.}
\label{fig8}
\end{figure*}

Following the procedure described in Ref. \cite{walther}, we also calculated the local
surface slope $\theta$ using the AFM images.
The local slope images corresponding to the height images shown in Fig. \ref{fig3} are
displayed in Fig. \ref{fig8}.
They clearly show that larger values for $\theta$ are obtained in intergranular regions,
while small $\theta$ are related to flat areas near the top of Cu$_2$O grains.

The distributions of local slopes are shown in Figs. \ref{fig8}e and \ref{fig8}f
for Cu$_2$O/n-Si and Cu$_2$O/Ni films, respectively. The position $\bar{\theta}$ of the
peaks of those distributions are displayed in Fig. \ref{fig8}g as a function of thickness.
For films grown on n-Si, the distribution significantly changes as the thickness increases,
with an increase in $\bar{\theta}$. This is typical of anomalous scaling~\cite{lopez,ramasco}
and is qualitatively consistent with the local roughness increase shown in Fig. \ref{fig5}a.
On the other hand, for films grown on Ni/n-Si, average local slopes decrease with the
thickness, saturating in the thicker films. This is consistent with normal scaling.
Comparison of local slope distributions with those of theoretical models is not possible
because there are significant thickness effects and the present concept of local slope
differs from the definitions in integrated growth equations or in lattice models.

The MH scaling was formerly observed in electrochemically deposited Prussian Blue
films~\cite{alamini}. In that case, the anomalous roughening was generated by time increasing
adsorption rates.

\subsection{Extended analysis of MH scaling}
\label{extended}

The exact solution of the MH equation~\cite{krug,amar1993} allows an extension of the
interpretation of the experimental data, particularly those of films grown on n-Si substrates.

The MH roughness $W(l,H)$ is expected to scale similarly to the height-height
correlation function; in terms of box size $l$ and time $t$ we have~\cite{krug}
\begin{equation}
W^2\left( l,t\right) \sim \frac{D}{K} l^2 \ln{\left( \xi / l \right)} F\left( l/\xi \right) ,
\label{eqWMH}
\end{equation}
where $F$ is a scaling function [$F(x) \approx const$ for small $x$] and
\begin{equation}
\xi = {\left( 2K t \right)}^{1/4} .
\label{eqxiMH}
\end{equation}
Recall that $H$ is proportional to the time $t$.

The long time (large thickness) scaling of the correlation length gives exponent $z$
near the MH value ($1/z \sim 1/4$). Thus, Eq. \ref{eqxiMH} indicates that $K$ is
approximately time-independent. The coefficient $K$ in Eq. \ref{eqMH} represents the
relation between surface diffusion coefficients and local surface geometry of a given
material. This relation is not expected to change in the course of the (constant temperature)
deposition.

\begin{figure}[h]
\includegraphics[width=8.0cm]{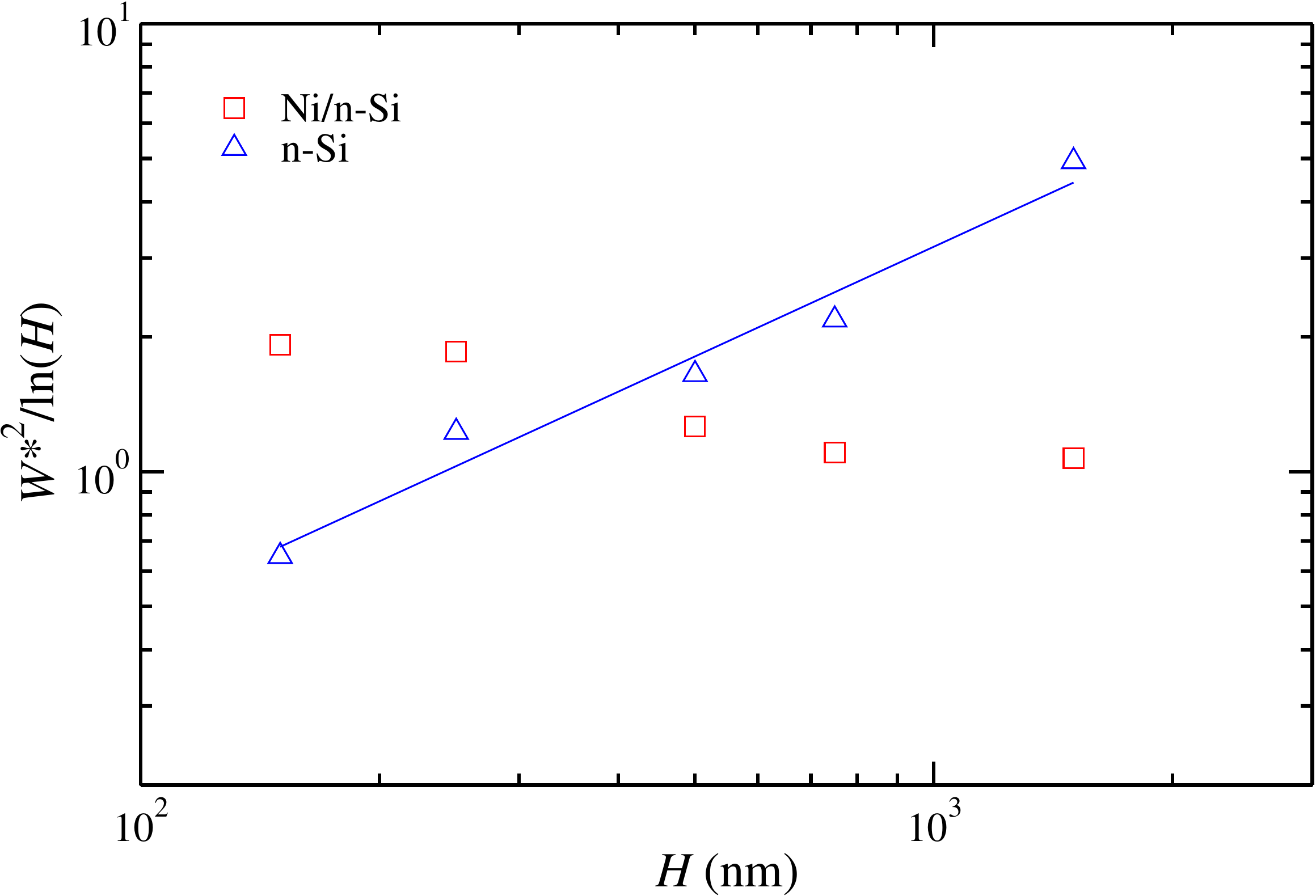}
\caption{(Color online) ${W^*}^2 /\ln{H}$ as a function of the thickness $H$ for films
grown on n-Si (triangles) and on Ni/n-Si (squares).
The solid line, with slope $0.8$, is a least squares fit of the large thickness data for
films grown on n-Si.}
\label{fig9}
\end{figure}

For $l \ll \xi$, Eq. \ref{eqWMH} suggests that the local roughness $W^*$ (measured for
fixed $l^* = 40$ nm; Sec. \ref{ResDynamic}) scales as ${W^*}^2 \sim \frac{D}{K} \ln{H}$.
Figure \ref{fig9} shows ${W^*}^2 /\ln{H}$ versus $H$ for both films.
The approximately constant value of ${W^*}^2 / \ln{H}$ for the films grown on Ni/n-Si
indicates that $D$ is also time-independent in that case.
However, the scaling in films grown on n-Si and the time-independence of $K$ gives
$D \sim t^{0.8}$.
Since $D$ is the amplitude of time and spatial fluctuations of deposition rate,
this means that those fluctuations increase in time in the films grown on n-Si.
This leads to the anomalous scaling in those films.

The above increase of $D$ is sufficiently slow, thus the original solution of the MH equation
can be consistently used with the time-dependent form of that parameter~\cite{krug,amar1993}.
This feature was already illustrated in other continuous and lattice models with time
dependent couplings~\cite{anomalouscompetitive,pradas2006} showing anomalous scaling.

\section{Discussion and conclusion}
\label{Secconclusion}

The scaling of distributions of local heights, maximal heights, and roughness provide
strong evidence that the roughening in electrodeposited Cu$_2$O films is governed by
the MH equation (\ref{eqMH}).
The noise amplitude increases in time in the equation associated to the films grown
on n-Si substrates, corresponding to the scaling anomaly.
This analysis clearly advance over the calculation of
scaling exponents, which however are also compatible with the MH ones.

The MH equation was proposed for roughening in molecular beam epitaxy~\cite{barabasi},
in which a crystalline film slowly grows by random adsorption of atoms/molecules from
vapor and surface adatoms diffuse and preferentially bind to high coordination
(low energy) sites.
Thus, in the electrochemical deposition of Cu$_2$O films, those results indicate that
this interplay of surface diffusion and deposition noise is also the main ingredient
for roughening under the present growth conditions.

The MH equation considers that adsorption rates are spatially uniform and do not depend
on time. This suggests that electrochemical conditions do not determine the main
dynamic mechanisms for roughening, although they probably affect the values of the
parameters in the associated MH equation. Since the same electrochemical conditions were
used for growth of films on both substrates, we are led to the conclusion that their
different structures are consequence of the substrate-film interaction.
The different interactions in different substrates determine the island nucleation and
growth and has consequences in the thickest films, despite the universal MH dynamics

The connections between the microscopic dynamics and the structure of the growing films
can be derived from our results, as follows.

In systems with diffusion-dominated dynamics, molecules at a surface grain randomly move,
with rates depending on local energy barriers.
At first approximation, those barriers depend on coordination numbers and
increase at step edges. For an adatom or ad-molecule to go down a step,
the additional activation energy, known as Ehrlich-Schwoebel (ES)
barrier~\cite{ehrlich,schwoebel}, reduces the downward flux near the step edge,
favoring nucleation of new atomic terraces over the ones previously
formed~\cite{evansreview}.
The effect of ES barriers is enhanced at grain boundaries, constraining the mass flux
between the grains if height differences are large.
These features are observed in films of several materials grown by vapor methods.
The above results strongly suggest to extend this interpretation to
Cu$_2$O electrodeposition.

Figure \ref{fig3}a shows that films grown on n-Si substrates have an initially small
average grain size, but with a broad size distribution. This is confirmed
quantitatively by the small values of the correlation length in Fig. \ref{fig5}d
($\xi\approx 80$ nm in $75$ nm thick films).
Thus, the islands nucleating at that substrate also had large size fluctuations.
This may be a consequence of large spatial fluctuations of the adsorption rate
during the first stages of the growth.

When an extended film is formed, the local adsorption rates may become uniform,
particularly if the dominant crystallographic orientation does not change,
as shown in Sec. \ref{ResCharact}; see also inset of Fig. \ref{fig1}b.
However, larger grains have larger numbers of terraces and steps, consequently
the mass flux from their tops to their boundaries is small.
With restricted mass exchange in grain boundaries, the large grains increase
in width and cover the smaller ones.
From a coarse grained point of view, the fluctuations in the adsorption rate
continue to increase, which qualitatively explains the time increase of the
parameter $D$ in the corresponding hydrodynamic model (the MH equation).

Figure \ref{fig8}e shows that the shape of the slope distribution is the same
while its average value moves to the right during the film growth.
This means that local slopes increase inside the grains and at their boundaries,
consistently with the above interpretation.

On the other hand, Fig. \ref{fig3}c shows that films grown on Ni/n-Si substrates
have larger initial grains with a much narrower size distribution (see also insets
of Figs. \ref{fig5}d and \ref{fig5}f). This is related to small fluctuations in
local adsorption rates, a feature that is maintained during the growth and leads
to normal roughening.

AFM images also reveal that the grain top surfaces are similar in the films
grown in both substrates, thus the local (microscopic) adsorption rates at those
surfaces are expected to be the same.
In growth on n-Si substrates, the difference in the net growth rate of individual
grains have to be explained by the restricted mass flux across grain boundaries,
not by intra-grain features.
If some grains with large slopes are initially formed, they may grow faster at the
expenses of the neighboring ones.
This establishes a connection between the V-shaped grains and the anomalous roughening.

Unusual scaling features are frequently found in diffusion-dominated
growth in the presence of step energy barriers~\cite{evansreview}.
For instance, a small barrier in Fe/Fe(100) epitaxy leads to mound formation at
the film surface, with skewed height distribution and $\beta = 0.2$~\cite{barteltPRL1995},
which are characteristics of the non-linear molecular beam epitaxy
equation~\cite{villain,laidassarma}.
Moreover, mound steepening in Ag/Ag(100) epitaxy at room temperature occurs with an
effective exponent $\beta$ much larger than $0.5$ in thicknesses ranging from 100 to
1000 monolayers~\cite{caspersenPRB2002} ($\beta = 0.5$ is a feature of completely
uncorrelated growth~\cite{barabasi}).

We conclude that, although control of electrochemical conditions is essential to enable
growth of Cu$_2$O films and to determine the crystallographic orientation,
our results show that large scale surface features are determined by physical
properties of the material and the substrate, with consequences on the internal film
structure. Roughening is governed by the interplay of deposition and ad-species surface
diffusion, similarly to vapor deposition, with no significant influence of inhomogeneous
mass flux from the solution (if it exists).
However, substrate-film interactions have a strong effect on island nucleation
and growth, thus the structure of the tickest films is drastically affected by
these initial conditions, despite the universal (MH) roughening dynamics.

\acknowledgments

The authors acknowledge financial support from CAPES, CNPq (NAMITEC and Nanoinstrumenta��o),
FAPESC, FAPEMIG and FAPERJ (Brazilian agencies). LCME, LabMat and LDRX laboratories of
the Universidade Federal de Santa Catarina for the use of the TEM, Raman spectroscopy
and XRD facilities.



\begin{references}

\bibitem{Delatorre2006} R. G. Delatorre, M. L. Munford, R. Zandonay, V. C. Zoldan, A. A. Pasa,  W. Schwarzacher, M. S. Meruvia, and I. A. H\"ummelgen, Appl. Phys. Lett. \textbf{88}, 233504 (2006).

\bibitem{Joseph2008} D. P. Joseph, T. P. David, S. P. Raja, and C. Venkateswaran, Mater. Charact. \textbf{59}, 1137 (2008).

\bibitem{Morales-Guio2014} C. G. Morales-Guio, S. D. Tilley, H. Vrubel, M. Gr\"atzel, and X. Hu, Nat. Commun. \textbf{5}, 3059 (2014).

\bibitem{Deng2013} M.-J. Deng, C.-Z. Song,  P.-J. Ho, C.-C. Wang, J.-M. Chen, and K.-T. Lu, Phys. Chem. Chem. Phys. \textbf{15}, 7479 (2013).

\bibitem{Liu2011} X.-W. Liu, Langmuir \textbf{27}, 9100 (2011).

\bibitem{Oba2005} F. Oba, F. Ernst, Y. Yu, R. Liu, H. M. Kothari, and J. A. Switzer, J. Am. Ceram. Soc. \textbf{88}, 253 (2005).

\bibitem{Chen2009} A. Chen, H. Long, X. Li, Y. Li, G. Yang, and P. Lu, Vacuum \textbf{83}, 927 (2009).

\bibitem{Deuermeier2011} J. Deuermeier, J. Gassmann, J. Br\"otz, and A. Klein, J. Appl. Phys. \textbf{109}, 113704 (2011).

\bibitem{Mittiga2006} A. Mittiga, E. Salza, F. Sarto, M. Tucci, and R. Vasanthi, Appl. Phys. Lett. \textbf{88}, 163502 (2006).

\bibitem{Zang2013} Z. Zang, A. Nakamura, and J. Temmyo, Opt. Express \textbf{21}, 11448 (2013).

\bibitem{Golden1996} T. D. Golden, M. G. Shumsky, Y. Zhou, R. A. VanderWerf, R. A. Van Leeuwen, and J. A. Switzer, Chem. Mater. \textbf{8}, 2499 (1996).

\bibitem{Switzer2002} J. A. Switzer, H. M. Kothari, and E. W. Bohannan, J. Phys. Chem. B \textbf{106}, 4027 (2002).

\bibitem{Brandt2014} I. S. Brandt, C. A. Martins, V. C. Zoldan, A. D. C. Viegas, J. H. D. da Silva, and A. A. Pasa, Thin Solid Films \textbf{562}, 144 (2014).

\bibitem{Bijani2009} S. Bijani, L. Mart\'inez, M. Gab\'as, E. A. Dalchiele, and J.R. Ramos-Barrado, J. Phys. Chem. C \textbf{113}, 19482 (2009).

\bibitem{Han2009} K. Han, and M. Tao, Sol. Energy Mater. Sol. Cells \textbf{93}, 153 (2009).

\bibitem{Gamburg2011} Y. D. Gamburg, and G. Zangari, \textit{Theory and Practice of Metal Electrodeposition} (Springer, 2011), p. 378.

\bibitem{Paracchino2012} A. Paracchino, J. C. Brauer, J.-E. Moser, E. Thimsen, and M. Graetzel, J. Phys. Chem. C \textbf{116}, 7341 (2012).

\bibitem{silvaSS2005} R. C. da Silva, A. A. Pasa, J. J. Mallett, W. Schwarzacher, Surf. Sci. \textbf{576}, 212 (2005).

\bibitem{barabasi} A.-L. Barabasi and H. E. Stanley, \textit{Fractal Concepts in Surface Growth} (Cambridge University Press, Cambridge, England, 1995).

\bibitem{krug} J. Krug, Adv. Phys. \textbf{46}, 139 (1997).

\bibitem{evansreview} J. W. Evans, P. A. Thiel, and M. C. Bartelt, Surf. Sci. Rep. {\bf 61}, 1 (2006).

\bibitem{Hua2011} Q. Hua, D. Shang, W. Zhang, K. Chen, S. Chang, Y. Ma, Z. Jiang, J. Yang, and W. Huang, Langmuir \textbf{27}, 665 (2011).

\bibitem{Somasundaram2007} S. Somasundaram, C. R. N. Chenthamarakshan, N. R. de Tacconi, and K. Rajeshwar, Int. J. Hydrogen Energy \textbf{32}, 4661 (2007).

\bibitem{Pallecchi2010} I. Pallecchi, L. Pellegrino, N. Banerjee, M. Cantoni, A. Gadaleta, A. S. Siri, and D. Marr\'e, Phys. Rev. B \textbf{81}, 165311 (2010).

\bibitem{Delatorre2009} R. G. Delatorre, V. Stenger, V. C. Zoldan, D. L. da Silva, S. G. dos Santos, A. D. C. Viegas, and A. A. Pasa, ECS Trans. \textbf{23}, 77 (2009).

\bibitem{Meyer2012} B. K. Meyer, A. Polity, D. Reppin, M. Becker, P. Hering, P. J. Klar, T. Sander, C. Reindl, J. Benz, M. Eickhoff, et al., Phys. Status Solidi \textbf{249}, 1487 (2012).

\bibitem{lopez} J. M. L\'opez, Phys. Rev. Lett. \textbf{83}, 4594 (1999).

\bibitem{ramasco} J. J. Ramasco, J. M. L\'opez, and M. A. Rodr\'{\i}guez, Phys. Rev. Lett. {\bf 84}, 2199 (2000).

\bibitem{huo} S. Huo, and W. Schwarzacher, Phys. Rev. Lett. \textbf{86}, 256 (2001).

\bibitem{lafouresse} M. C. Lafouresse, P. J. Heard, and W. Schwarzacher, Phys. Rev. Lett. \textbf{98}, 236101 (2007).

\bibitem{tiago2007} T. J. Oliveira and F. D. A. A. Reis, J. Appl. Phys. \textbf{101}, 063507 (2007).

\bibitem{tiago2011} T. J. Oliveira and F. D. A. Aarao Reis, Phys. Rev. E \textbf{83}, 041608 (2011).

\bibitem{kpz} M. Kardar, G. Parisi, and Y.-C. Zhang, Phys. Rev. Lett. \textbf{56}, 889 (1986).

\bibitem{fabio2004} F. D. A. Aarao Reis, Phys. Rev. E {\bf 69}, 021610 (2004).

\bibitem{kelling} J. Kelling and G. \'Odor, Phys. Rev. E \textbf{84}, 061150 (2011).

\bibitem{mh} W. W. Mullins, J. Appl. Phys. \textbf{28}, 333 (1957); C. Herring, in \textit{The Physics of Powder Metallurgy}, edited by W. E. Kingston (McGraw-Hill, New York, 1951).

\bibitem{Almeida2014} R. A. L. Almeida, S. O. Ferreira, T. J. Oliveira, and F. D. A. Aarao Reis, Phys. Rev. B \textbf{89}, 045309 (2014).

\bibitem{siniscalco} D. Siniscalco, M. Edely, J.-F. Bardeau, and N. Delorme, Langmuir \textbf{29}, 717 (2013).

\bibitem{zoldan2005} V. C. Zoldan, D. M. Kirkwood, G. Zangari, M. L. Munford, W. Figueiredo, and A. A. Pasa, Microsc. Microanal. \textbf{11}, 154 (2005).

\bibitem{colaiori} F. Colaiori and M.A. Moore, Phys. Rev. Lett. \textbf{86}, 3946 (2001).

\bibitem{healy2014} T. Halpin-Healy and G. Palasantzas, Europhys. Lett. \textbf{105}, 50001 (2014).

\bibitem{Almeida2015} R. A. L. Almeida, S. O. Ferreira, I. R. B. Ribeiro, and T. J. Oliveira, Europhys. Lett. \textbf{109}, 46003 (2015).

\bibitem{halpin2012} T. Halpin-Healy, Phys. Rev. Lett. \textbf{109}, 170602 (2012); Phys. Rev. E \textbf{88}, 042118 (2013).

\bibitem{tiagorapid2013} T. J. Oliveira, S. G. Alves and S. C. Ferreira, Phys. Rev. E \textbf{87}, 040102(R) (2013).

\bibitem{lcm} J. M. Kim and S. Das Sarma, Phys. Rev. Lett. \textbf{72}, 2903 (1994).

\bibitem{racz1994} G. Foltin, K. Oerding, Z. R\'acz, R. L. Workman, and R. K. P. Zia, Phys. Rev. E \textbf{50}, R639 (1994).

\bibitem{antal2002} T. Antal, M. Droz, G. Gy\"orgyi, and Z. R\'acz, Phys. Rev. E \textbf{65}, 046140 (2002).

\bibitem{walther} L. Liu and W. Schwarzacher, Electrochem. Commun. \textbf{29}, 52 (2013).

\bibitem{alamini} M. F. Alamini, R. C. da Silva, V. C. Zoldan, E. A. Isoppo, U. P. R. Filho, F. D. A. A. Reis, A. N. Klein, and A. A. Pasa, Electrochem. Commun. \textbf{13}, 1455 (2011).

\bibitem{amar1993} J. G. Amar, P.-M. Lam, and F. Family, Phys. Rev. E \textbf{47}, 3242 (1993).

\bibitem{anomalouscompetitive} F. D. A. Aar\~ao Reis, Phys. Rev. E {\bf 84}, 031604 (2011).

\bibitem{pradas2006} M. Pradas and A. Hern\'andez-Machado, Phys. Rev. E {\bf 74}, 041608 (2006).

\bibitem{schwoebel} R. L. Schwoebel and E. J. Shipsey, J. Appl. Phys. \textbf{37}, 3682 (1966).

\bibitem{ehrlich} G. Ehrlich and F. G. Hudda, J.Chem. Phys. \textbf{44}, 1039 (1966).

\bibitem{barteltPRL1995} M. C. Bartelt and J. W. Evans, Phys. Rev. Lett. {\bf 75}, 4250 (1995).

\bibitem{villain} J. Villain, J. Phys. I {\bf 1}, 19  (1991).

\bibitem{laidassarma} Z.-W. Lai and S. Das Sarma, Phys. Rev. Lett. {\bf 66}, 2348 (1991).

\bibitem{caspersenPRB2002} K. J. Caspersen, A. R. Layson, C. R. Stoldt, V. Fournee, P. A. Thiel, and J. W. Evans, Phys. Rev. B \textbf{65}, 193407 (2002).









\end{references}
\end{document}